\documentclass[12pt]{article}
\usepackage{epsfig}
\begin{document}
\hfill AS-TEXONO/01-04 \\
\hspace*{1cm} \hfill \today

\begin{center}
{\Large  \bf
Nuclear Recoil Measurement in CsI(Tl) Crystal\\
for Cold Dark Matter Detection} \\
\vskip 0.5cm
\large
M.Z.~Wang$^{a}$,
Q.~Yue$^{b}$,
J.R.~Deng$^{c}$,
W.P.~Lai$^{d,e}$,
H.B.~Li$^{a,d}$,
J.~Li$^{b}$,\\
Y.~Liu$^{b}$,
B.J.~Qi$^{c}$,
X.C.~Ruan$^{c}$,
C.H.~Tang$^{a}$,
H.Q.~Tang$^{c}$,\\
H.T.~Wong$^{d,}$\footnote{Corresponding~author:
Email:~htwong@phys.sinica.edu.tw;
Tel:+886-2-2789-9682;
FAX:+886-2-2788-9828.},
S.C.~Wu$^{a}$,
B.~Xin$^{c}$,
Z.Y.~Zhou$^{c}$

\normalsize
\vskip .2cm

\begin{flushleft}
{$^{a}$\rm
Department of Physics, National Taiwan University, Taipei, Taiwan.\\}
{$^{b}$\rm 
Institute of High Energy Physics, Beijing, China.\\}
{$^{c}$\rm 
Department of Nuclear Physics, Institute of Atomic Energy, Beijing, China\\}
{$^{d}$\rm 
Institute of Physics, Academia Sinica, Taipei, Taiwan.\\}
{$^{e}$\rm
Department of Management Information Systems,
Chung Kuo Institute of Technology, \\
\hspace*{1cm} Taipei, Taiwan.\\}
\end{flushleft}

\clearpage

\end{center}
\vskip 0.5cm
\begin{abstract}

There are recent interests with
CsI(Tl) scintillating crystals for Dark Matter
experiments.
The scattering signatures by neutrons on a CsI(Tl) 
detector were studied
using a neutron beam generated by a 13~MV Tandem accelerator.
The energy spectra of 
nuclear recoils from 7 keV to 132 keV were measured,
and their quenching factors for scintillating light yield 
were derived.
The data represents the first
confirmation of the Optical Model 
predictions on neutron elastic scatterings with a
direct measurement on the nuclear recoils of heavy nuclei.
The pulse shape discrimination techniques to differentiate
nuclear recoils from $\gamma$-background were
studied.  Internal consistencies were obtained among
the different methods of light yield measurements.
The projected capabilities for Cold Dark Matter
searches with CsI(Tl) crystals are presented. 
\end{abstract}

\begin{flushleft}
\end{flushleft}

\vfill

\newpage

\section{Introduction}

The detection of Dark Matter and the studies of their
properties~\cite{dmref} are of fundamental importance in
particle physics and cosmology.
The Weakly Interacting Massive Particles (WIMPs)
are good candidates for ``Cold'' Dark Matter,
and their experimental searches
have gathered a lot of interests in recent years. 
The most promising avenue is to detect the
nuclear recoil signatures due to elastic scatterings
of WIMPs on the target isotopes.
The typical energy depositions are
only of the order of 10~keV, imposing
big experimental challenges in terms
of the detection of weak signals as well as 
background control at low energy close to detection
threshold.
A wide spectrum of experimental techniques is 
being pursued~\cite{dmexpt}.

The DAMA experiment observed an annual modulation
of nuclear recoil events~\cite{DAMA} with 
NaI(Tl) scintillating crystal detectors,
which can be interpreted as positive evidence
of WIMPs due to the difference of the relative
velocities of the Earth from the Halo sea within
the year. 
This, however, contradicts the limit
from the CDMS experiment~\cite{CDMS} 
based on  cryogenic technique.

There is still much room for new detector concept 
to thoroughly test the DAMA parameter space, and
to push the sensitivities further. It would be
of great interest if the sensitivities of 
WIMP searches can probe the level predicted by
the various Super-Symmetry models.

There are potential merits of using CsI(Tl) scintillating
crystals~\cite{csichar} for WIMP search and other low-energy
low-background experiments~\cite{prospects,ksexpt}.
An experiment towards 200~kg of CsI(Tl) crystal
scintillators to study low energy neutrino
interactions at the Kuo-Sheng power reactor is
being pursued~\cite{ksexpt}, while
the adaptation of the crystal
for Dark Matter searches
are the focus of several on-going 
projects~\cite{csidmfrance,csidmuk,csidmkorea}.

The high-A content of the CsI enhances
the sensitivities for the spin-independent interactions
(which depends on $\rm{ A^2}$)
between the WIMPs and the target,
relative to most other candidate target isotopes.
The high-Z composition allows
a compact design and provides large suppression of
background due to ambient radioactivity if
a three dimensional fiducial volume definition
can be realized.
Both $^{133}$Cs and $^{127}$I are 100\% in their
respective isotopic abundance. Being close in
their mass numbers, the response to nuclear recoil
from the interactions with WIMPs would be similar,
allowing simpler interpretation of experimental
signatures.

As a detector, the crystal  
has large light yield, low energy
threshold and with pulse shape
discrimination characteristics for $\gamma$/$\alpha$
separation. Its characteristics of being only
slightly hygroscopic implies that it
can be machined easily and does not
require hermetic seal (that is, passive materials)
in a large detector system. 
In addition, large (40~tons) electromagnetic
calorimeter systems~\cite{bfactories}
have been constructed and
made operational in high energy physics experiments,
making this technology affordable and
realistic to scale up.
Considering all the associated costs, the price
of CsI(Tl) is in fact less than that for NaI(Tl).
In  order to produce positive and definite evidence
of the WIMPs, 
an accurate measurement of the annual modulation
(where the maximal effects are only 7\%)
would be necessary such that
the availability of large target mass 
is a very desirable feature.

One of the key issues to realize a Dark Matter search
experiment with CsI(Tl) crystal scintillator
is the studies of the experimental signatures 
of nuclear recoils due to WIMP-nuclei elastic
scatterings. Nuclear recoils produce 
large charge density (dE/dx)
such that the scintillating light yield is ``quenched''
and the timing profile of pulse shape is different
relative to the same energy deposition by minimum ionizing
particles~\cite{scinbasic}. 

These signatures are the same as the nuclear recoil
events produced by elastic scattering of neutrons
on nuclei. A measurement was performed with
the neutron facility at the 13~MV Tandem accelerator
at the China Institute of Atomic Energy at Beijing.
The results of the quenching factors measurement
and of pulse shape discrimination studies are
reported in this article.
They extend the work from the other recent measurements
on CsI(Tl)~\cite{csidmfrance,csidmuk},
as well as on other scintillating crystals
such as NaI(Tl) and CaF$_2$(Eu)~\cite{nbeamnai}.

\section{Experimental Setup and Procedures}
\label{sect::setup}

\begin{figure}
\epsfig{file=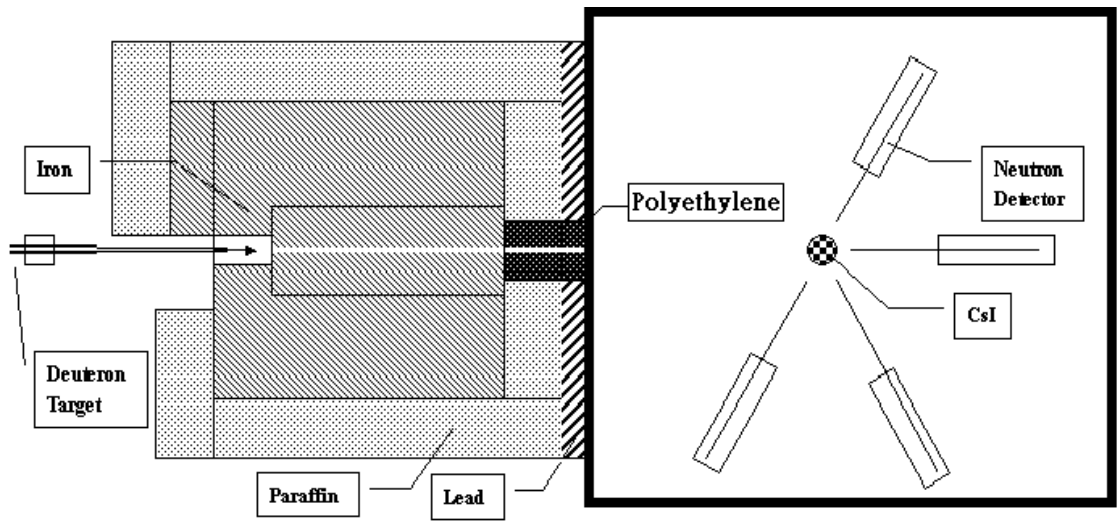,width=15cm}
\caption{ 
Schematic diagram of the experimental set-up.
}
\label{setup}
\end{figure}

The experiment was performed at HI-13 tandem accelerator  at the
China Institute of Atomic Energy (CIAE) in Beijing.
A schematic diagram is shown in Figure~\ref{setup}. 
A pulsed deuteron beam at 5.6~MeV 
interacted with a deuterium gas target in
a cell at a pressure of 6~bar and the dimensions
of 1 cm in diameter and 3 cm in length.
Neutrons at 8~MeV kinetic energy
were produced at zero degree.
The CsI(Tl) sample was located 2.02 meters away from 
the deuterium target. 
Neutrons at zero degree were selected by a 
32 mm by 35 mm collimator of length 1.2~m.
The collimator were surrounded by shielding
materials like iron, paraffin,
lead and polyethylene to reduce background.
During this data taking, 
the repetition rate of the pulsed beam was set at 
4~MHz and the average beam current was at 1~$\mu$A.

The scattering target, which also functioned as a detector,
was a CsI(Tl) 
crystal scintillator\footnote{Producer: Unique Crystal, Beijing}
with diameter 3~cm and length 3~cm
wrapped with teflon sheets, aluminium foil and black plastic tape.
To minimize supporting structures, the detector was hung
at the correct position by a piece of string.
The readout was achieved by a 29~mm diameter
photo-multiplier tube (PMT)\footnote{CR110, Hamamatsu Photonics, China} 
The PMT signals passed through an amplifier and shaper and
were digitized by a 20~MHz, 8-bit, Flash
Analog-to-Digital Convertor (FADC)~\cite{electronics}
developed for the Kuo-Sheng reactor neutrino experiment~\cite{ksexpt}.
The trigger was provided from a time of flight (TOF)
system  to tag the elastically scattered neutrons.
The $\gamma$-response of the CsI(Tl) detector
was calibrated with an LED pulser operated at
single photo-electron intensity, as
well as standard 
$^{109}$Cd and $^{133}$Ba
$\gamma$-sources. 
The calibration is
4 photo-electrons  per keV of electron-equivalence energy.

Neutron tag was provided by 
liquid scintillator\footnote{Co-261, ST-451} detectors 
105 mm in diameter and 50 mm in length 
equipped with PMT\footnote{Philips XP-2041} readout. 
The liquid scintillator provides  
pulse shape discrimination (PSD) capabilities
for n-$\gamma$ differentiation.
The PSD was achieved by 
commercial electronics\footnote{CANBERRA 2160A}.
The neutron tag signal was used
to define the ``START'' timing for the TOF system.
The pulsed deuteron ``pick-off'' signal
provided by the accelerator was
delayed and used as ``STOP'' for the TOF. 
The neutron detectors were placed at a distance of 
0.5 to 2 m from the CsI(Tl) target at different angles,
to provide optimized TOF resolution to differentiate events
due to neutron elastic scattering
off hydrogen atoms or heavier nuclei  in the
wrapping materials.    

The FADC recorded data continuously in a circular
buffer of 4k size.
A START-STOP sequence from the TOF system
initiated a trigger which stopped the FADC  
digitization after 25~$\mu$s.
The pulse shape from the CsI(Tl) crystal 
was recorded for a pre-trigger and post-trigger periods of 
5~$\mu$s and 25~$\mu$s, respectively.
With the timing settings at data taking,
the CsI(Tl) pulse started at $-$1.25~$\mu$s
if it was due to nuclear recoils of heavy nuclei.
This timing information provides a powerful
means of the rejection of background due to
proton recoils and pile-up events.

\begin{figure}
\epsfig{file=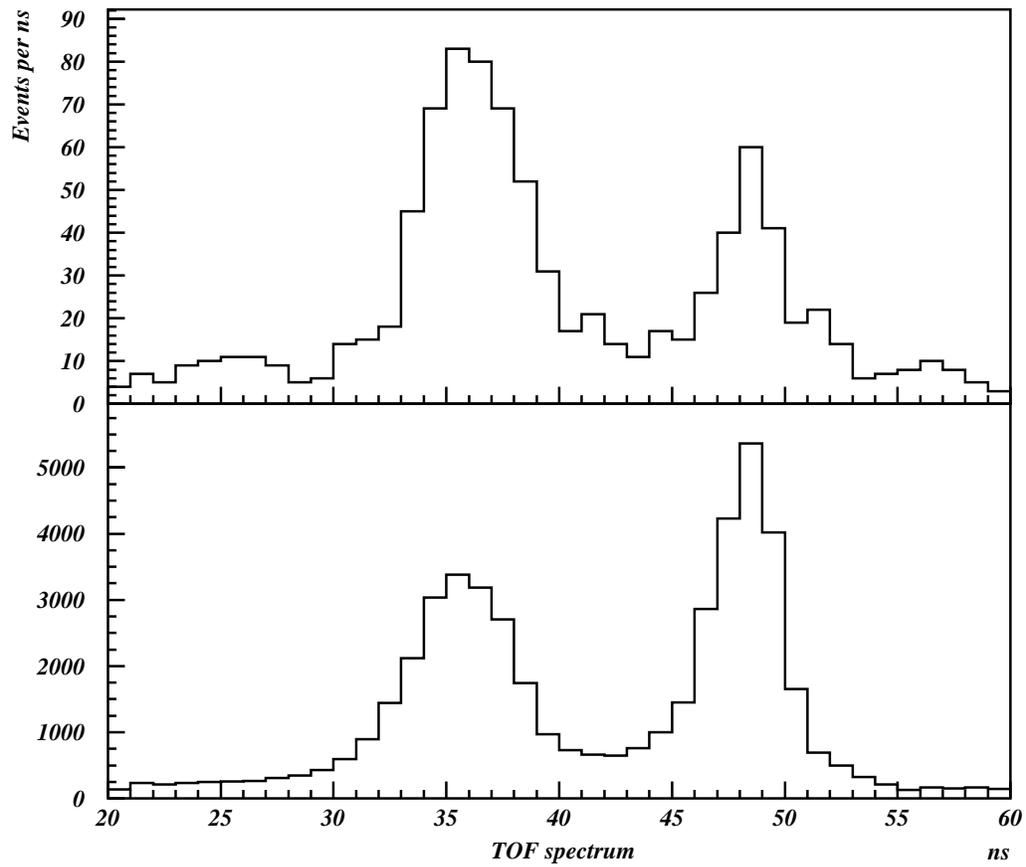,width=15cm}
\caption{ The TOF spectra at 50$^{\circ}$ scattering angle for
(a)with wrapping material only, (b) with CsI(Tl) target.
}
\label{TOF}
\end{figure}

Data taking at each scattering angle                               
was complemented by a measurement
of the TOF background spectrum  
with {\it empty} target 
which was made of  
the CsI(Tl) wrapping materials only. 
Displayed in 
Figure~\ref{TOF} is the TOF spectra at 50$^{\circ}$ with 
empty and complete target.
The lower and upper peaks are due to neutron elastic scattering off
protons and the 
heavier nuclei such as carbon, aluminium,
cesium, and iodine, respectively. 
The flat background is due to random coincidence.
It can be seen that with the CsI(Tl) target
in place, the upper TOF peak for neutron scattering 
off heavier nuclei becomes enhanced. 
The difference between the two TOF spectra is the
nuclear recoils spectra from Cs and I.
The proton-recoil events tagged
by the low TOF peak
provides a means to evaluate  
the signal yield and 
to confirm the distance between neutron detector and
CsI target. 

\begin{figure}
\epsfig{file=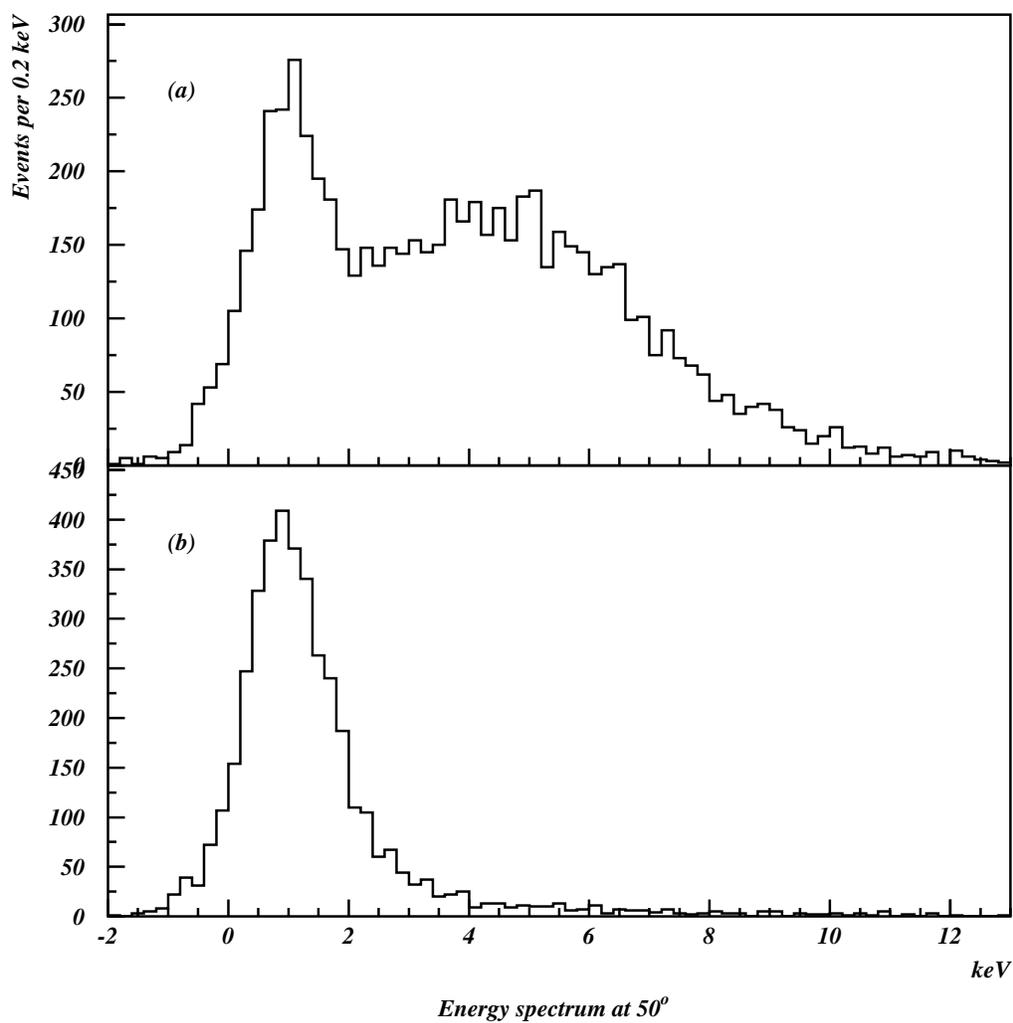,width=15cm}
\caption{ The energy spectra at 50$^{\circ}$ scattering angle for
(a)events with high TOF values, (b) events with low TOF values.
}
\label{E50}
\end{figure}

The reconstructed energy is defined as 
the pedestal-subtracted integrated area of
the FADC pulse over 2.5~$\mu$s starting from 
the t$_0$ of FADC 
(1.25~$\mu$s before TOF trigger in this data taking).
This restricted  time window helps
to minimize the pile up effect from
accidental low energy $\gamma$'s 
which are abundant along the beam line.   
Events are categorized into two groups by the central value 
between the two TOF peaks and their corresponding 
nuclear recoil energy spectra 
are compared. The energy spectra at 50$^{\circ}$ are shown 
in Figure~\ref{E50}a and \ref{E50}b for events with 
high TOF values and
low TOF values, respectively. 

The energy spectrum tagged by the low TOF trigger
is displayed in Figure~\ref{E50}b.
Since the trigger is due to proton recoils
from the wrapping materials, there should
be no correlated signals from the CsI(Tl) target. 
The energy spectrum
which shows a distinguished peak
with long tail on the high-energy side
is due to events  
from accidental $\gamma$'s arriving
within the integration time-window. 

The detector response for CsI  nuclear recoil events
was obtained by subtracting Figure~\ref{E50}b
from Figure~\ref{E50}a with proper normalizations,
where the factors 
are derived by normalizing
the low TOF peaks from proton recoils
between
the full and empty target spectra.

\section{Measurement Results}
\label{sect::statsub}

\begin{figure}
\epsfig{file=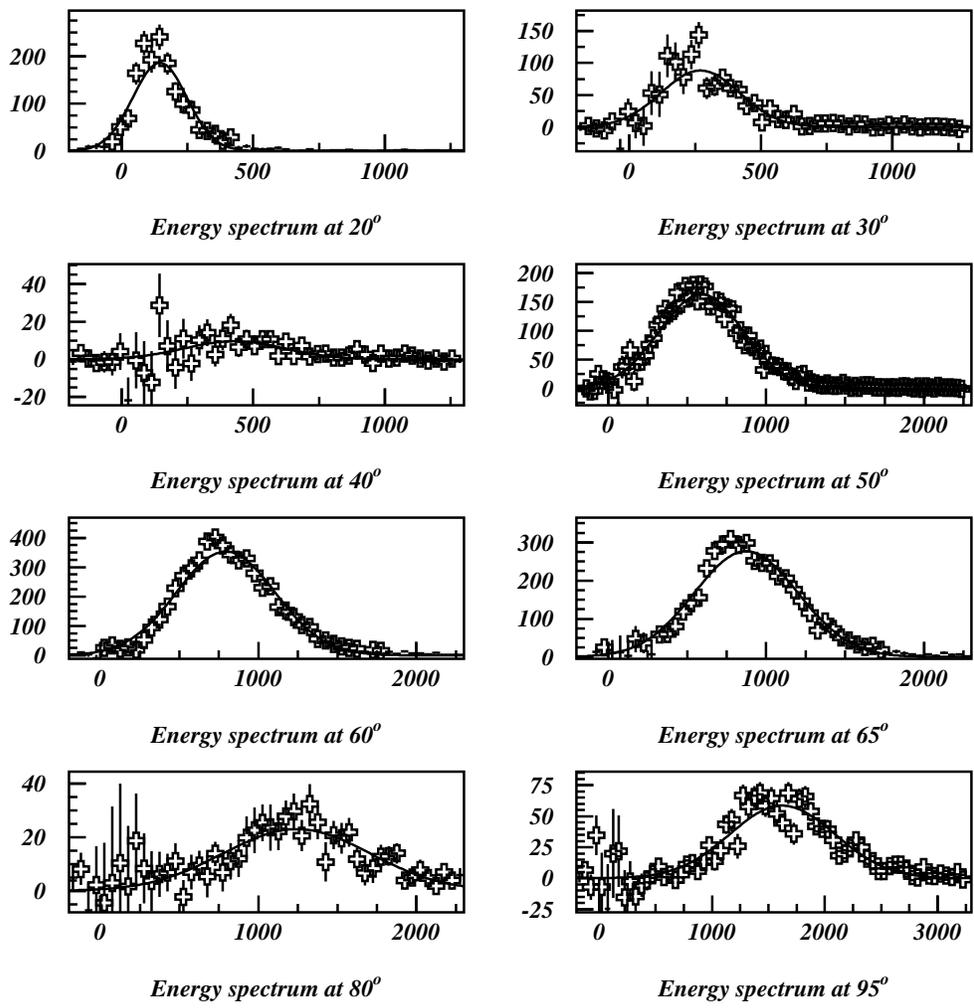,width=15cm}
\caption{ The observed energy spectra with background
          subtracted at different scattering angles.
}
\label{Energy}
\end{figure}

The resulting CsI recoil spectra 
obtained by the ``statistical background subtraction'' scheme
discussed in Section~\ref{sect::setup}
are shown in Figure~\ref{Energy} for eight different scattering
angles. 
It can be seen that this procedure is valid
since the background levels away from the peaks
after subtraction are consistent with zero.

The nuclear recoil energy (T) is related to the
neutron scattering angle ($\theta$) by a simple
kinematical formula:
\begin{equation}
{\rm
T ~ = ~ {2A \over (1+A)^2} ~ (1-cos\theta) ~ E_n ~~~~ ,
}
\end{equation}
where $\rm{A}$ is the target atomic mass and 
$\rm{E_n}$ is the incident neutron
energy. The neutron scattering angle can be considered
to be the same in this case in both
the laboratory and center-of-mass systems.
Comparisons of the nuclear recoil spectra with those
due to calibration sources allow the evaluation of the 
quenching factors, defined as the light yield from
nuclear recoils versus that from $\gamma$ sources.

To enhance the signal-to-noise ratio
in the large background environment,
the nuclear recoil pulses are integrated
for a restricted 2.5~$\mu$s (or 50 FADC time bins),
as compared to the
full time window 12.5 $\mu$s (or 250 FADC time bins)
for 100\% light collection in the
$\gamma$-sources calibration measurements.
A correction factor of 1.15 is applied
to the nuclear recoils light yield
to account for this partial summation.
This factor was derived by averaging and comparing
a large number of events due to nuclear recoils
and $^{109}$Cd $\gamma$-source, as depicted in
Figure~\ref{Pulse}.
The estimated uncertainty 
is 10\%, based on studies
with different
integration time window from 1.5 $\mu$s to 12.5 $\mu$s.

The deviation from the 
central value of nuclear recoil energy 
due to finite detector size 
is at most 2\% for different scattering angles. 
This is checked by 
calculating the average recoil energy with proper weights of event 
angular  distribution and detector acceptance. A shift of detector at
1 mm scale has negligible effect on the central value of nuclear 
recoil energy. 
The beam current was stable at the 1\% level
during data taking, as indicated by the stable data taking rate (typical 
value is about 4 Hz) as well as readings from beam monitor
counters. 
The main systematic uncertainties originate from the
analysis procedures, and amount to 16\% at the lowest
energy point at 20$^{\circ}$.
The statistical and systematic uncertainties 
are then combined in quadrature to provide
the total uncertainties for each data point. 

\begin{figure}
{\bf (a)}\\
\centerline{
\epsfig{file=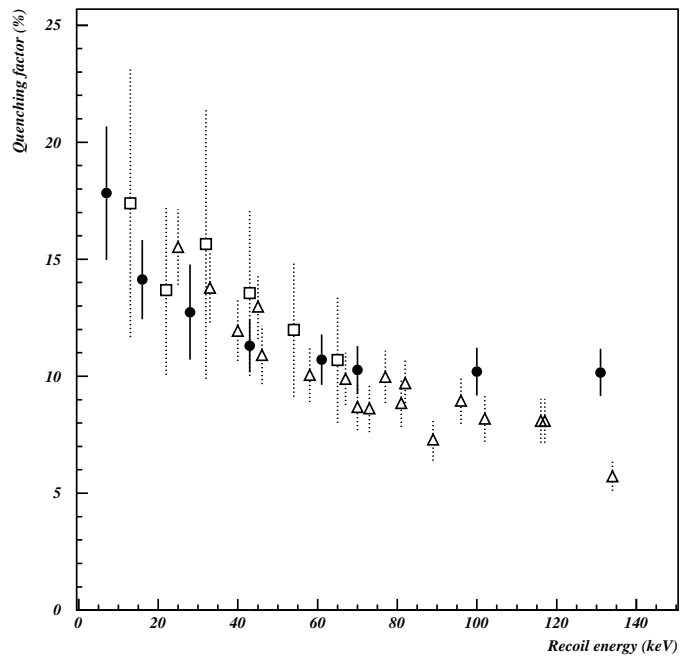,width=10cm}\\
}
{\bf (b)}\\
\centerline{
\epsfig{file=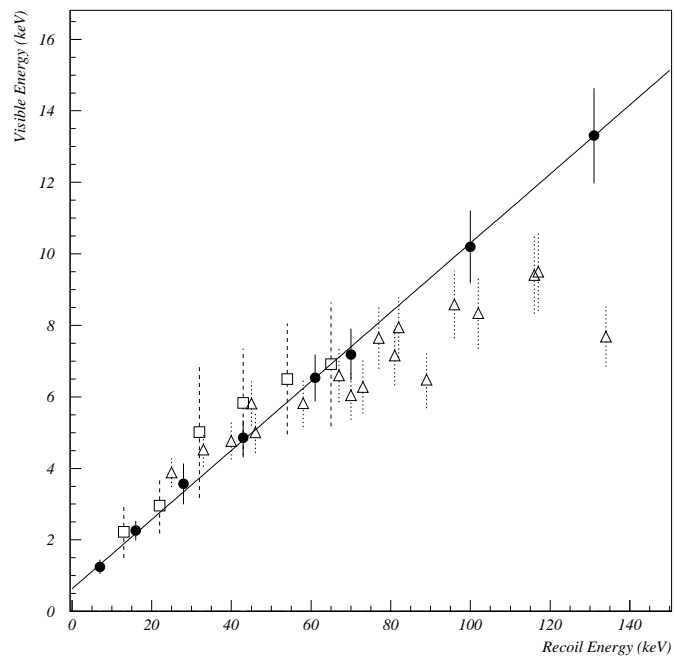,width=10cm}
}
\caption{
(a) The quenching factors and 
(b) electron-equivalence light yield versus recoil energy
measured in this work shown as black circles. 
The solid line is a linear fit to the data.
Open triangles and open squares are data
from Ref.~\cite{csidmfrance} and  
Ref.~\cite{csidmuk}, respectively. 
}
\label{Quenching}
\end{figure}

The results at different scattering angles are summarized
in Table \ref{tabresults}. 
The measured quenching factors are depicted
in Figure~\ref{Quenching}a, while the variation
of electron-equivalence light yield (the ``visible'' energy)
versus recoil energies are shown in Figure~\ref{Quenching}b.
The other measurements~\cite{csidmfrance,csidmuk}
are overlaid for comparison.
There is a clear trend of less quenching towards low energy points.
This measurement achieved a lower threshold and with improved
uncertainty compared to Ref.~\cite{csidmuk}.
The quenching factors are of the similar range to
those for NaI(Tl)~\cite{naiqf},
which are typically 25\% for Na and 8\% for I.
It can be seen that the light yield is higher
for interactions with the high-A isotopes in CsI(Tl),
providing big advantages to probing the spin-independent
interactions.
In addition, Figure~\ref{Quenching}b depicts
a linear relationship between the electron-equivalence 
light yield and the nuclear recoil energy, as indicated
by a linear fit our data set. 
Data from Ref.~\cite{csidmfrance} shows
more quenching and deviates from the linear
regime at high energy.
This is different from the highly non-linear response
due to threshold effects
observed in liquid scintillator~\cite{qfliqscin}
at the same range of recoil energy.

The total number of observed signal events
are given by the area under the Gaussian fits in Figure~\ref{Energy}.
To derive the differential cross sections,
the data taking time, 
the acceptance (proportional to $r^{-2}$) and the
efficiency correction 
factors have to be evaluated.
The efficiency factors are due to  
the TOF selection criteria
and background subtraction scheme. 
The efficiency is close to 100\%
for the well-separated TOF peaks, but
is only 42\%
for the worst cases 
where the TOF peaks overlap.
The contributions from inelastic scattering
leading to the low excited levels for  $^{133}$Cs and $^{127}$I
are expected to be less than 0.1\% relative to the
elastic processes.

\begin{figure}
\epsfig{file=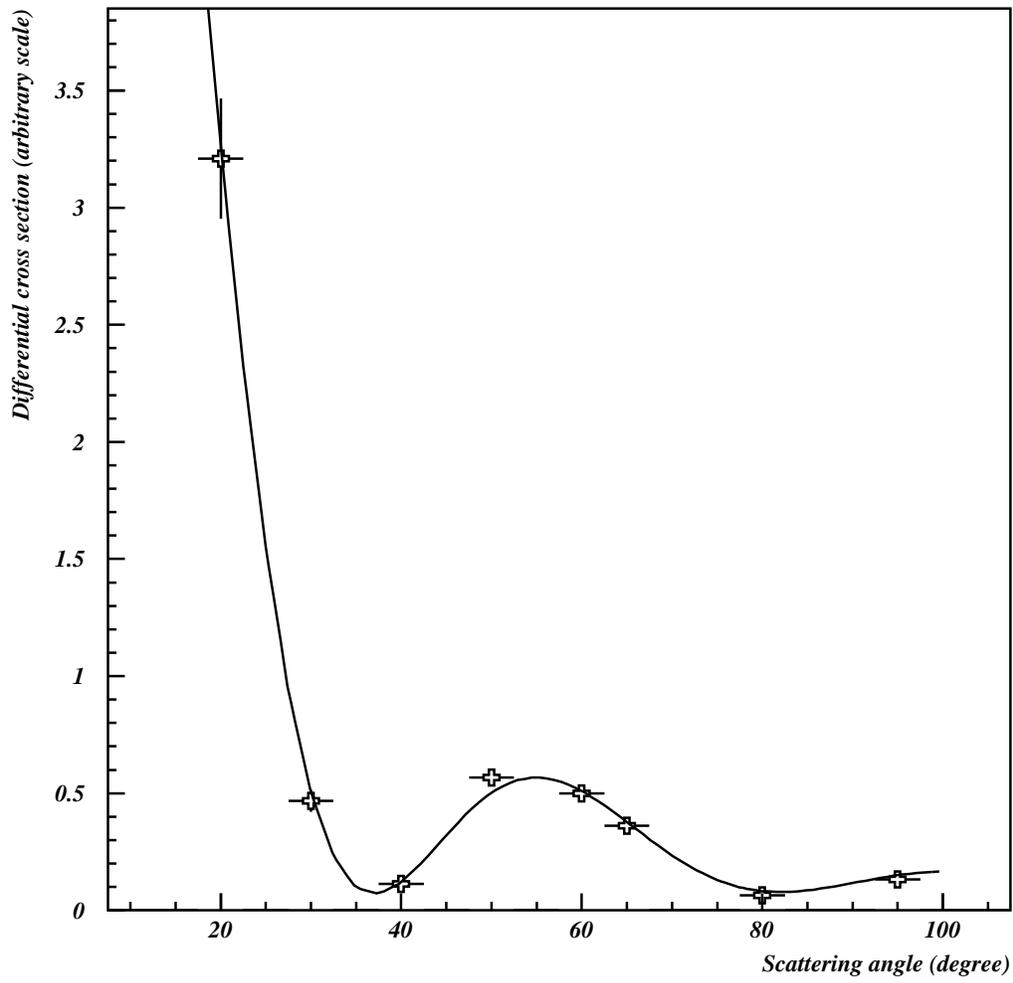,width=15cm}
\caption{
The differential cross sections 
determined from nuclear
recoils in the target.
}
\label{Xect}
\end{figure}

After all the correction factors 
are taken into account, 
the measured angular distribution
for neutron elastic scatterings
off Cs and I is displayed in
Figure~\ref{Xect}. 
The shape of the solid curve is the evaluated neutron elastic scattering
cross sections on $^{133}$Cs and $^{127}$I 
from ENDF/B-VI~library~\cite{nndc}
which comes from the Optical Model calculation.
The absolute normalization is from a best-fit value.
There is excellent agreement
between the relative recoil angular distribution
from this measurement 
and the predictions from the Optical Model.
In particular, this is a kinematical
regime where the size of the nuclei
is comparable to the wavelength of the incident neutrons,
and the diffraction pattern is  well-reproduced by the data.

The measurements demonstrate the validity of
the nuclear recoil event selection procedures
at all angles.
In addition, 
the results in Figure~\ref{Xect}
confirms the Optical Model predictions
on neutron elastic scatterings
from a direct measurement 
of nuclear recoils on heavy nuclei (instead of from the
scattered final-state neutrons in transmission-type
experiments) $-$ that is, they show that neutrons do
elastically scatter off from the nuclei but
not via some anomalous processes.

\section{Pulse Shape Discrimination}
\label{sect::psd}

\begin{figure}
\epsfig{file=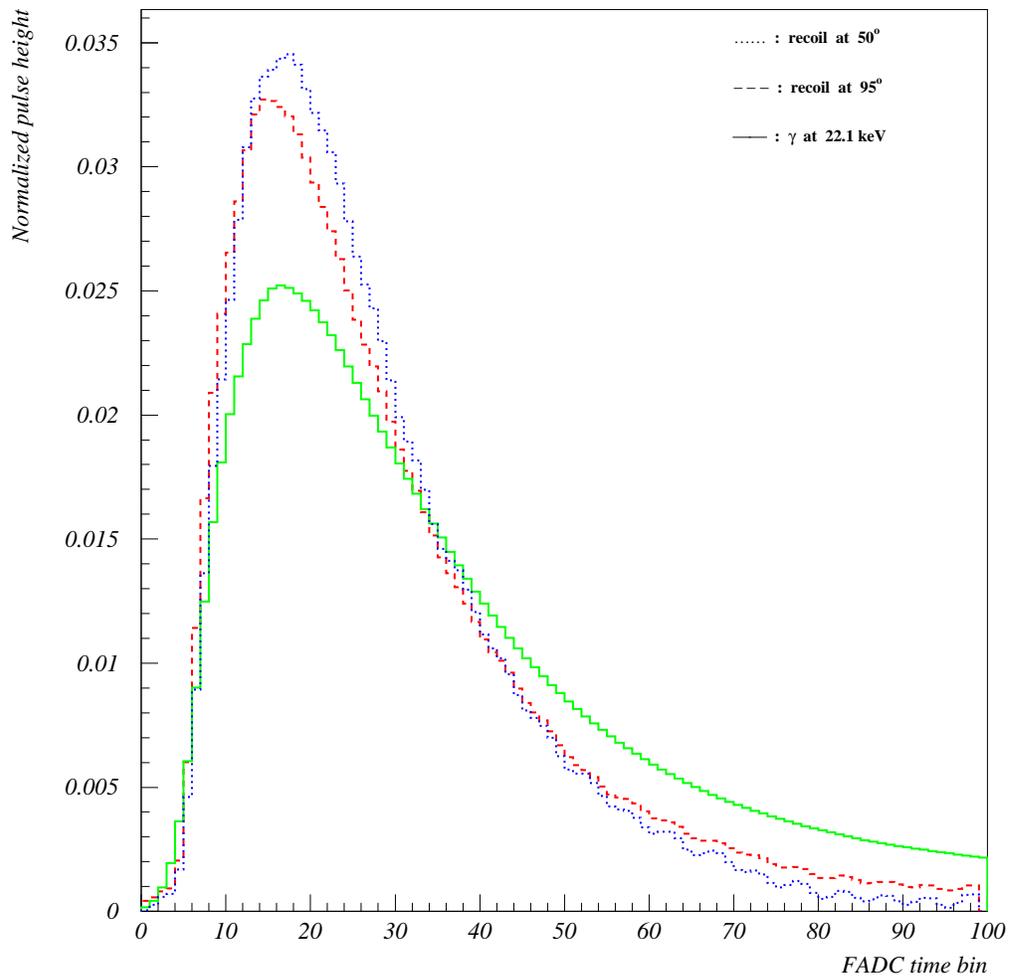,width=15cm}
\caption{
The normalized pulse shape in the first 100 FADC samplings
at 50~ns per bin.
Solid historgram is due to 22.1 keV X-rays, while
the two dashed histograms are from recoils
at 50$^{\circ}$ and 95$^{\circ}$.
}
\label{Pulse}
\end{figure}

The nuclear recoil measurements provide in addition
an excellent data set for the study of PSD  at
small photo-electron level near detection threshold.
Depicted in Figure~\ref{Pulse} are
the measured nuclear recoils
at 50$^{\circ}$ and 95$^{\circ}$,
as well as the $\gamma$-pulse due to the
22.1~keV line from $^{109}$Cd, 
averaged over many events and normalized to the
same light collecition.
It can be seen that heavily-ionizing events 
such as those due to $\alpha$-particles and nuclear
recoils have
{\it faster} decays than those from $\gamma$'s $-$
opposite to the response in
liquid scintillator~\cite{scinbasic}.

The neutron induced nuclear recoil
pulse shapes are similar between
50$^{\circ}$ and 95$^{\circ}$ scattering
angles, corresponding to 
4.9~keV and 13.4~keV electron-equivalence
light yield, respectively.
The shape is also similar to that of 
the $\alpha$ source data~\cite{proto} 
taken at high energy. 
At smaller scattering angles, the
accidental low-energy $\gamma$-background
becomes important such that the intrinsic
nuclear recoil pulse shape cannot be
derived.
In the PSD studies, the
recoil pulse shapes for these data set
were taken to be
the same as those measured at 50$^{\circ}$.
The similarity of the pulse shape between
50$^{\circ}$ and 50$^{\circ}$ justifies this
assumption.

\begin{figure}
\epsfig{file=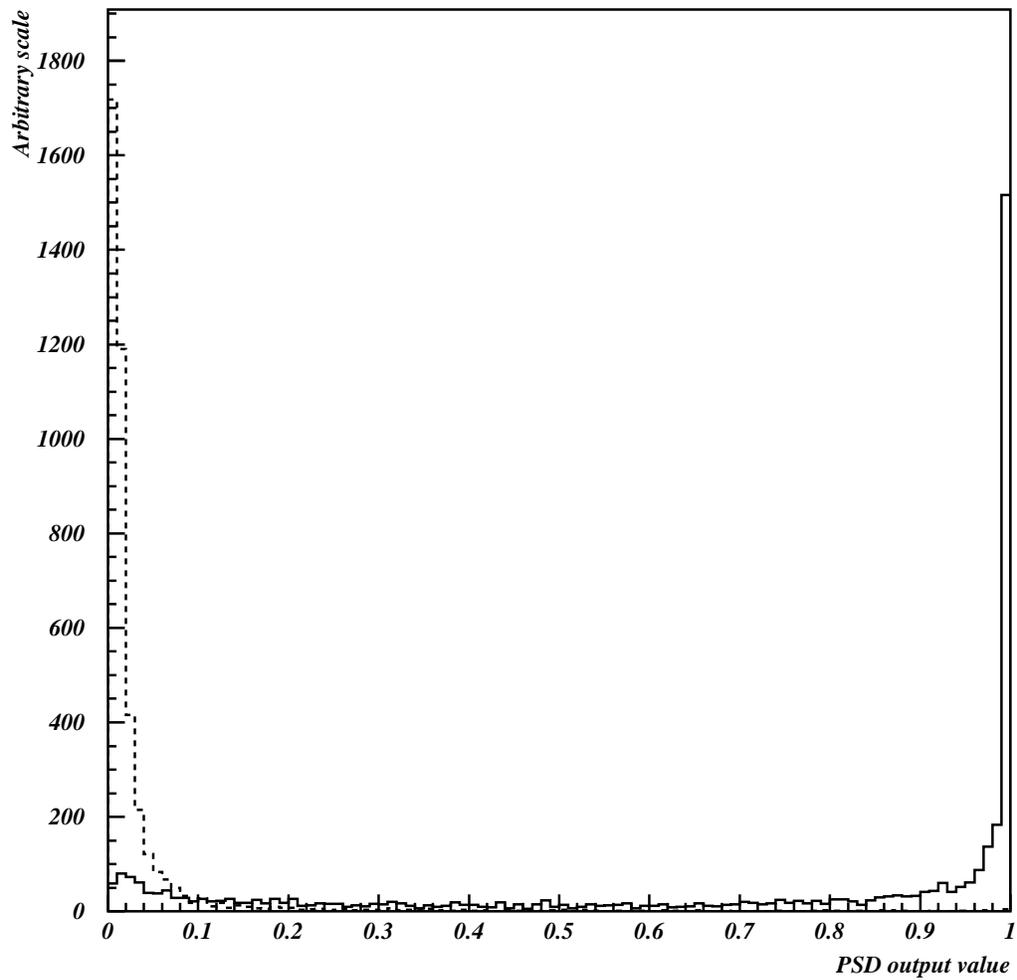,width=15cm}
\caption{
The neural network output of 4000 nuclear recoil events (solid histogram) 
and 4000 $\gamma$ events (dashed histogram). The nuclear recoil energy
is 43~keV and the $\gamma$ energy is 22~keV.  
}
\label{psd}
\end{figure}

Two independent approaches to obtain PSD were studied:
(1) neural network technique~\cite{neural} 
and (2) the ``classical'' method based on
mean-time plus partial charge.
For the neural network approach,
we used 4000 events each from the above two data sets as training
samples and additional 4000 events each as testing samples. 
The trained neural net has
20 input nodes (that is, 5 FADC time bins combined
to form one node in Figure~\ref{Pulse})
in 5 $\mu$s window 
with 10 hidden nodes. The input node value should be greater than
zero after
pedestal correction. A negative value is reset to zero and the sum of
the 20 input values are normalized to one.
The algorithm used is back propagation. 
The excellent separation power from neural net study
is depicted in Figure~\ref{psd}.
The trained neural net is applied to the complete 50$^{\circ}$ data set.
Good separation is obtained
between the 43~keV nuclear recoils and the low energy
$\gamma$ pulses.
A cut on neural output greater
than 0.9 selects almost all the CsI nuclear recoil events.
The number of events as well as the mean and RMS 
electron-equivalence energy
are consistent with those determined from the statistical
background subtraction method in Section~\ref{sect::statsub}. 
The PSD separation with neural network method
does not apply to the 20$^o$ data set. 
The neutron-beam environment gave rise to
large accidental background 
(mostly low energy $\gamma$-events) which
produce much stronger bias for
the low energy nuclear recoil signatures.
This would not be a problem in a realistic Dark
Matter experiment where the background rate would
be extremely low.

\begin{figure}
\epsfig{file=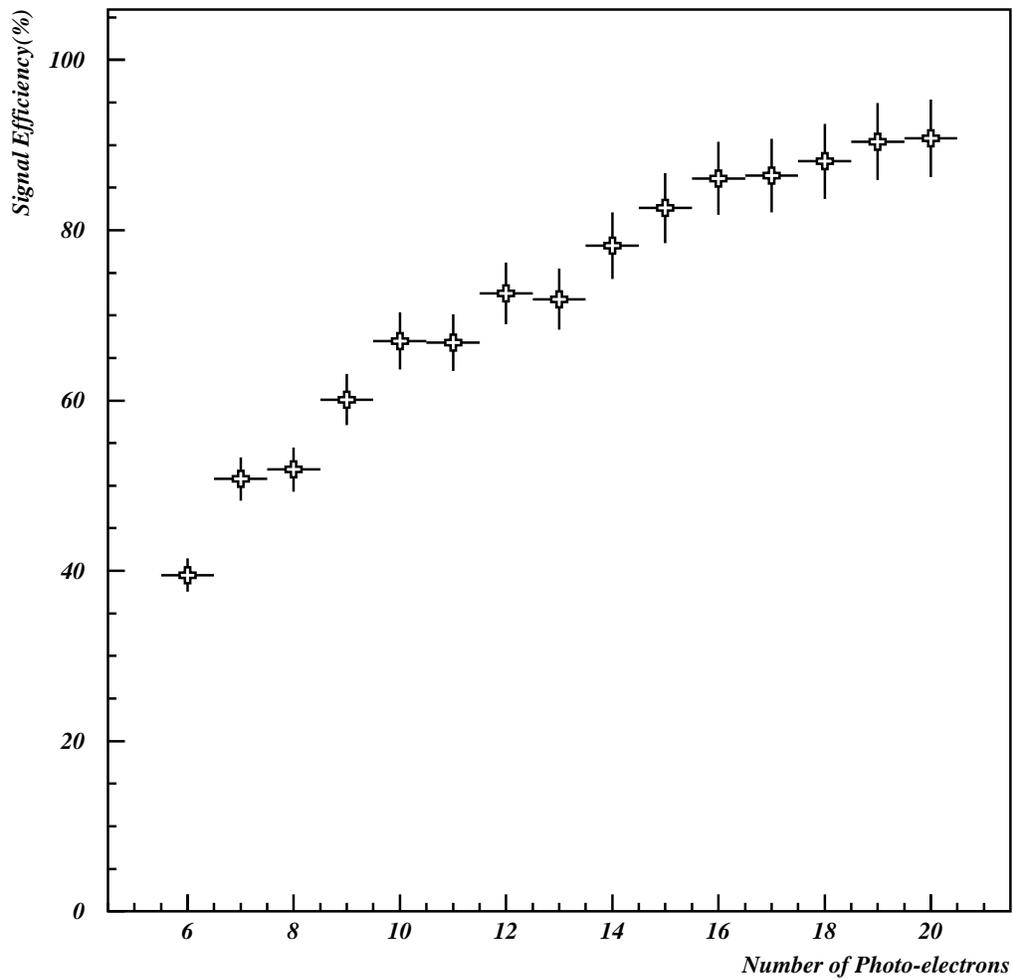,width=15cm}
\caption{
The PSD performance based on
neural net studies. The $\gamma$
rejection is fixed at 95\% and the result of signal efficiency is 
plotted against available number of photo-electrons.    
}
\label{photoelec}
\end{figure}

A large simulated sample was generated  
at different
photon-statistics situation 
with these recoil/$\gamma$ pulse shapes,
and the results are displayed in Figure~\ref{photoelec}. 
It can be seen that
a 95\% $\gamma$ rejection with
$>$90\% efficiency for pulses with
an average 20 photo-electrons can be achieved. 
It is projected that the performance
can be further improved in a low-background
quiet environment which allows a longer sampling window.
These are better than the PSD capabilities 
for NaI(Tl) detectors at the same light yield~\cite{csidmuk}.

\begin{figure}
\centerline{
\epsfig{file=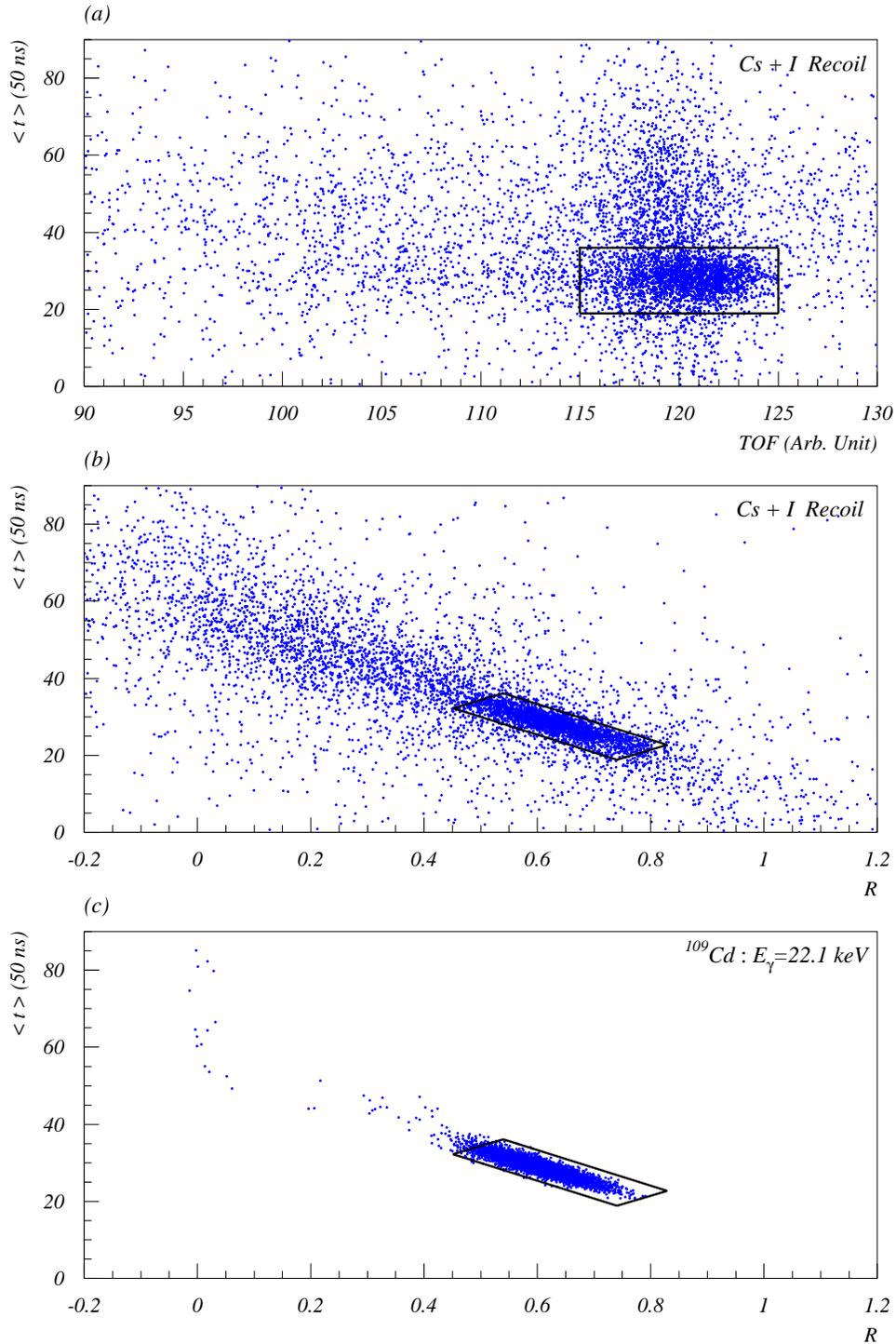,width=15cm}
}
\caption{
Scattered plots for
the $\rm{ \langle t \rangle }$ parameter versus the
(a) TOF values and 
(b) R parameters for the nuclear recoil events
at 95$^{\circ}$, 
and (c) for events due to
the 22~keV $\gamma$-rays from a $^{109}$Cd source.
The box region represents the CsI nuclear recoil events,
indicating clear seperations
from the accidental $\gamma$-background.
}
\label{psd2d}
\end{figure}

The alternative and complementary approach is one
based on mean-time and partial charge.
Two variables, the mean time $\rm{ \langle t \rangle }$
and the partial charge ratio R, are 
defined  for each pulse, such that
\begin{equation}
\rm{
\langle t \rangle ~ = ~
 { \sum\limits_{i=1}^{50} ~ ( A_i ~  t_i ) \over \sum\limits_{i=1}^{50} ~ A_i }
}
\end{equation}
and
\begin{equation}
\rm{
R ~ = ~ { \sum\limits_{i=1}^{30} ~ A_i \over \sum\limits_{i=1}^{50} ~ A_i } ~~ ,
}
\end{equation}
where $\rm{A_i}$ is the FADC amplitude at time bin $\rm{t_i}$.
The scattered plots for $\rm{\langle t \rangle }$ versus TOF as well
as $\rm{\langle t \rangle}$  versus R for the data
set at 95$^{\circ}$ are shown in Figures~\ref{psd2d}a\&b,
respectively. The electron-equivalence energy for
CsI recoils is 13.4~keV at this scattering angle.
There are clear separations between the
Cs and I nuclear recoil events, represented by
the box region, and the
accidental $\gamma$-background for events
triggered by the recoils 
of the wrapping materials (protons as well as other
heavy nuclei).
The cuts in $\rm{\langle t \rangle }$ versus R space
as defined in Figure~\ref{psd2d}b are subsequently
applied to the other data set at lower recoil energies
to select nuclear recoil events.

The events due to the 22~keV $\gamma$-rays from
a $^{109}$Cd source are displayed in Figure~\ref{psd2d}c.
The good overlap between the box and the events
justifies that the cuts are valid to select ``in-time'' events.
However, the separation between
nuclear recoils from the in-time $\gamma$-background
is not satisfactory. 
While the classical method is
sufficient for event identification in {\it this} 
measurement where the main background are
the accidental low-energy $\gamma$'s, it would
be inadequate for a Dark Matter experiment where
the $\gamma$-background also provide the trigger
and are therefore in-time.
In comparison with Figure~\ref{photoelec}, 
the neural network method
gives 95\% rejection of the in-time $\gamma$'s
at a signal efficiency of better than 90\% at $>$20
photo-electrons (that is, larger than 50$^\circ$ in this
data set). 
The neural network method, therefore,
will give superior performance in a
Dark Matter experiment compared to the
classical method.

\begin{figure}
\epsfig{file=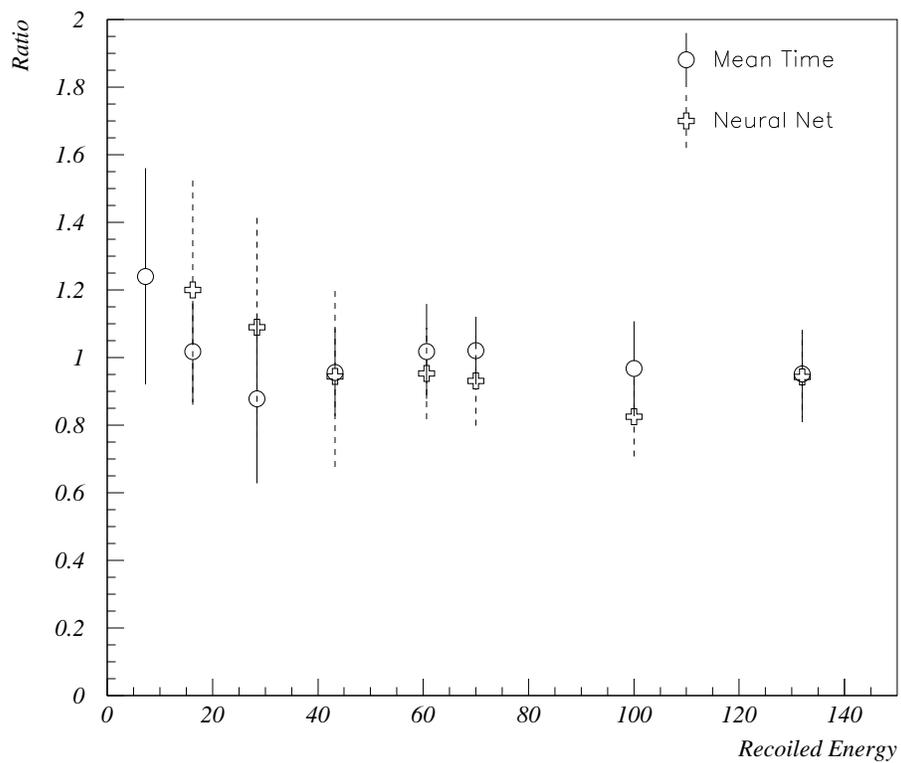,width=15cm}
\caption{
Ratio of electron-equivalence light yield derived from
the two PSD algorithms, 
labelled ``Mean Time'' and ``Neural Net'',
with the measurements of the 
statistical subtraction method. Consistent results
are obtained among them.
}
\label{yieldratio}
\end{figure}

The electron-equivalence light yield at different scattering
angles for the nuclear recoil events selected by either of 
the two PSD algorithms are measured. 
The comparison between the two PSD algorithms
with the subtraction method discussed in Section~\ref{sect::statsub}
is depicted in Figure~\ref{yieldratio}, indicating 
that consistent results are obtained. 
The deviations of the mean values
among these independent measurements
are also consistent with the estimated systematic
uncertainties. 

Therefore,
the robustness of the quenching factor results 
in Figures~\ref{Quenching}a\&b measured
by the statistical subtraction method 
discussed in Section~\ref{sect::statsub}
is further established.
This method involves a straight-forward
normalization and subtraction scheme, and 
is therefore expected to give more unbiased 
results compared
to the two PSD algorithms.

\section{Discussion and Summary}

In this article, we presented a new measurement on
the quenching factors of Cs and I in a CsI(Tl) crystal
scintillator based on statistical background subtraction. 
Lower threshold and improved accuracies
are achieved compared to previous measurements.
The measured differential cross-section 
of neutron scattering from I and Cs 
is in excellent
agreement with Optical Model derivations, and 
represents the first confirmation of 
the Optical Model on neutron
elastic scattering cross-sections
with a direct measurement of 
nuclear recoils from heavy nuclei.
The recoil differential cross sections and the quenching
factors are relevant to the studies of  
radiation damage in materials. 

Two complementary
pulse shaping discrimination techniques
at the low photo-statistics regime
were studied based on the nuclear recoil data.
It is shown that the neural network approach 
gives superior rejection power to in-time
$\gamma$-background compared to the
classical method based on mean-time
and partial charge. 
Both procedures are adequate
for differentiating accidental $\gamma$'s.
The electron-equivalence light yield derived
from both PSD methods are consistent with those
obtained by statistical subtraction. 
The internal consistencies of the light yield measurements
among the  three different algorithms  and also
with the Optical Model predictions of differential
cross section enhance the reliability and robustness
of the results.

By optimizing the detector geometry and using
green-extended photo-cathode,
a detector with several kg modular mass
and a light yield of a few photo-electron per keV 
can be realized~\cite{csidmkorea}. 
The background from ambient radioactivity
or intrinsic radio-purity is of course
crucial to all low-background experiments.
Their considerations for CsI(Tl) detectors
are discussed in Refs.~\cite{prospects,ksexpt,csibkg}.
Levels of better
than the $10^{-12}$~g/g level in
concentration for the $^{238}$U and
$^{232}$Th series have been demonstrated,
assuming secular equilibrium.
The potential problem of the internal $^{137}$Cs
contaminations can be overcome via selection
of clear ore materials and careful chemical
processing and purification treatment~\cite{csidmkorea}. 

\begin{figure}
\epsfig{file=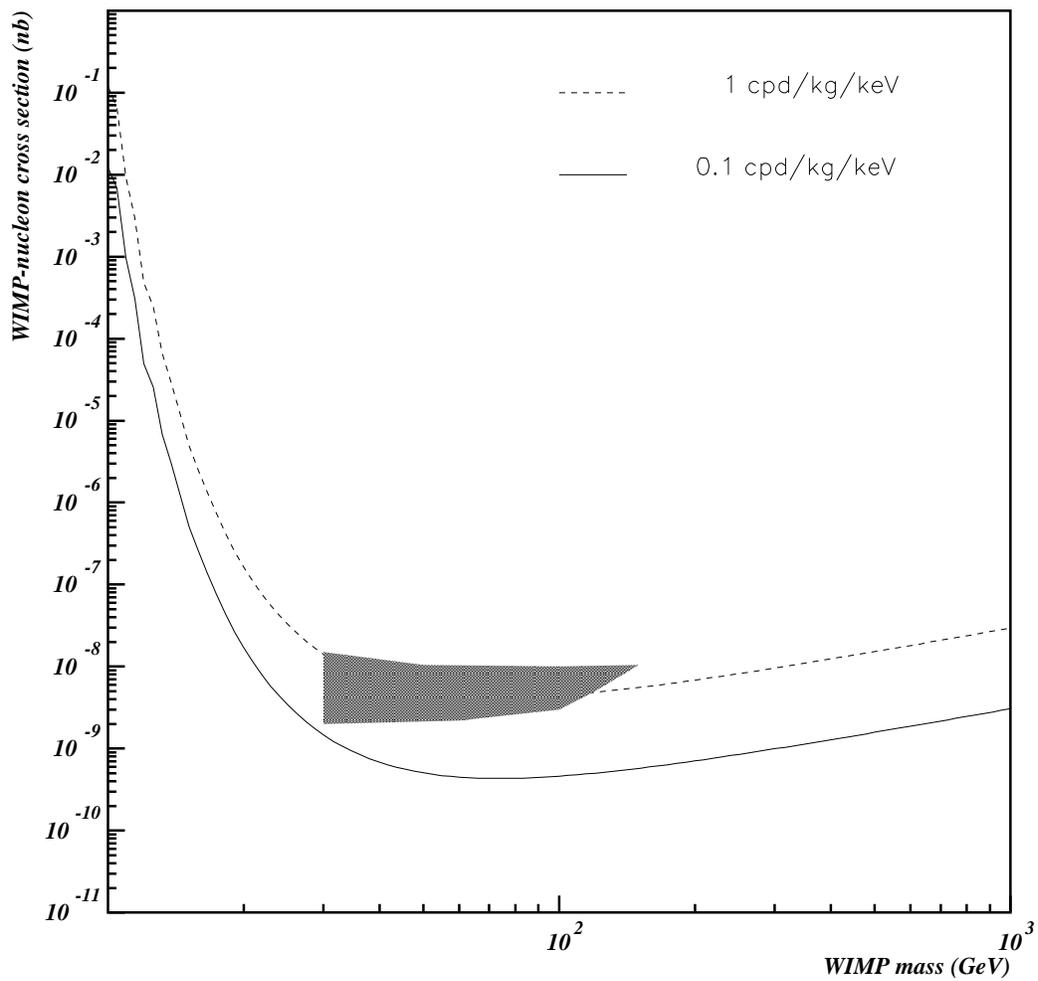,width=15cm}
\caption{
The sensitivity plot
of cross-section versus WIMP mass on a 1~ton CsI(Tl) detector,
with the assumptions that the detector threshold is set at
2 keV with a quenching factor of 0.13.
The two cases of background
counting rate at 1 and 0.1 counts per day (cpd) per kg per keV
are shown.
The DAMA allowed region is presented by the shaded area.
}
\label{Exclude}
\end{figure}

Displayed in Figure~\ref{Exclude} is 
the sensitivity plot for spin-independent 
WIMP interaction cross section per nucleon
with a target mass of 1~ton of CsI(Tl).
The sensitivities are 
based on the assumptions that
the detection threshold is 
at 2~keV (or about 16~keV recoil
energy) and still allows
enough photo-electron
statistics for background rejection with PSD,
and that
the observed background rate after applying
the PSD cuts
is at the same  as
or ten times better level than 
the NaI(Tl) experiments (1 and 0.1 per day per kg per keV,
respectively). 
It can be seen that
the DAMA allowed region, represented by
the shaded region in Figure~\ref{Exclude},
can be thoroughly probed.
Given that there is matured experience of scaling
up the CsI(Tl) detector to multi-ton systems,
there are potentials of further 
improvements in the sensitivities,
and in the case of positive results, performing
an accurate measurement of the annual modulation.


\section{Acknowledgements}

The authors would like to thank 
Drs. K.W. Ng, S.K. Kim and Y.D. Kim
for fruitful discussions and helpful comments,
and are grateful to 
the technical staff
from CIAE and IHEP for operating the accelerator.
This work was supported by contracts
CosPa~89-N-FA01-1-4-2 from the Ministry of Education, Taiwan,
NSC~89-2112-M-001-056 and NSC~90-2112-M-001-037
from the National Science Council, Taiwan,
and NSF19975050 from the
National Science Foundation, China.
The support of H.B.~Li is from  contracts
NSC~90-2112-M002-028 and MOE~89-N-FA01-1-0
under Prof.~P.W.Y.~Hwang.

\clearpage

\begin{table}[htb]
\large
\begin{center}
\begin{tabular}{|l||c|c|c|c|c|c|c|c|}
\hline
Angle ($^{\circ}$)&  20&  30&  40&  50&  60&  65&  80&  95\\ \hline \hline
Neutron Detector & 228& 133& 100&  89&  60&  63&  70&  68\\ 
\hspace*{0.2cm} Distance (cm) & & & & & & & & \\ \hline
Recoil Energy (keV)  &   7.3&  16.2&  28.4&  43.3&  60.6&  
70.0& 100& 132\\ \hline
Mean Light Output & 168 & 311 & 495 & 676 & 
 912 & 1004 & 1428 &  1866 \\
\hspace*{0.2cm} (FADC Unit) & & & & & & & & \\ \hline
Electron-Equivalence & 1.25 & 2.26 & 3.57 & 4.86 & 6.53 & 7.18 & 10.2 & 13.4 \\
\hspace*{0.2cm} Light Yield (keV) 
& $\pm$ & $\pm$ & $\pm$ & $\pm$ & $\pm$ & $\pm$ & $\pm$ & $\pm$ \\
&  0.20 &  0.27 & 0.57 & 0.49  & 0.65  & 0.72 & 1.0 & 1.3 \\ \hline
Quenching Factor & 0.17& 0.14& 0.13& 0.11& 0.11& 0.10 & 0.10 & 0.10 \\
& $\pm$ & $\pm$ & $\pm$ & $\pm$ & $\pm$ & $\pm$ & $\pm$ & $\pm$ \\
&  0.027 &  0.017 & 0.020 
& 0.011  & 0.011  & 0.010 & 0.010 & 0.010 \\ \hline
\end{tabular}
\end{center}
\caption{Measured results of the neutron elastic scatterings
from CsI(Tl) detector. Errors shown are combined systematic
and statistical uncertainties.}
\label{tabresults}
\end{table}

\normalsize

\clearpage

\clearpage

\clearpage

\clearpage

\clearpage

\clearpage

\clearpage

\clearpage

\clearpage

\clearpage

\clearpage

\clearpage

\end{document}